\documentclass{article}
\usepackage{spconf,amsmath,graphicx}
\usepackage{amsmath,graphicx}
\usepackage{algorithmic}
\usepackage[ruled,vlined]{algorithm2e}
\usepackage{changepage}
\usepackage{siunitx,booktabs}
\usepackage{enumerate,enumitem}
\usepackage{amssymb}
\newcommand{\trans}{^{\mathsf{T}}}
\usepackage{paralist}
\usepackage{multirow}

\def \PhiB {{\mathbf{\Phi}_\mathrm{B}}}
\def \Phitot {{\mathbf{\Phi}_\mathrm{tot}}}
\def \Phitote {{\mathbf{\Phi}_{\mathrm{tot,e}}}}
\def \Phitott {{\mathbf{\Phi}_{\mathrm{tot,t}}}}
\def \Phitotinv {{\mathbf{\Phi}_\mathrm{tot}^{-1}}}
\def \Phitoteinv {{\mathbf{\Phi}_{\mathrm{tot,e}}^{-1}}}

\title{Incorporating uncertainty from speaker embedding estimation to speaker verification}
\name{Qiongqiong Wang, Kong Aik Lee, Tianchi Liu}
\address{Institute for Infocomm Research (I$^2$R), A$^\star$STAR, Singapore}
\begin{document}
\ninept
\maketitle
\begin{abstract}
Speech utterances recorded under differing conditions exhibit varying degrees of confidence in their embedding estimates, i.e., uncertainty,  even if they are extracted using the same neural network.
This paper aims to incorporate the uncertainty estimate produced in the xi-vector network front-end with a probabilistic linear discriminant analysis (PLDA) back-end scoring for speaker verification.  
To achieve this we derive a posterior covariance matrix, which measures the uncertainty, from the frame-wise precisions to the embedding space. We propose a log-likelihood ratio function for the PLDA scoring with the uncertainty propagation. We also propose to replace the length normalization pre-processing technique with a length scaling technique for the application of uncertainty propagation in the back-end. 
Experimental results on the VoxCeleb-1, SITW test sets as well as a domain-mismatched CNCeleb1-E set show the effectiveness of the proposed techniques with $14.5\%$--41.3$\%$  EER reductions and $4.6\%$--$25.3\%$ minDCF reductions.

\end{abstract}
\section{Introduction}
\label{sec:intro}
Automatic speaker verification (ASV) is the process to verify whether a given speech utterance is from a specific speaker or not. Recent ASV systems have benefited from deep learning by replacing individual components \cite{Snyder18,Tang19,Chien17}, as well as the entire pipeline in an end-to-end manner \cite{Li17, Rohdin20}. Among these, it has been shown to be the most viable and effective to use DNNs in the front-end for discriminative speaker embedding extractions.
Speaker embeddings are fixed-length continuous-value representations of input sequences that contain acoustic feature vectors living in large, complex spaces. 
Using it simplifies ASV task because in the embedding space, probability distributions and geometric concepts are easily applicable. 
Therefore, speaker embeddings and a scoring back-end are often used together in ASV frameworks \cite{i4u19,matejka20,villalba20,nagrani20, lee20}.

Substantial amount of works have reported on network architectures that produce embedding vectors with better speaker representations \cite{Snyder18,Desplanques20, lee21,Liu22}.
Unlike the generative i-vector extraction model in which a posterior mean and precision are calculated simultaneously \cite{dehak11}, deep speaker embedding networks mostly only produce a points estimate without a guarantee or a measure of its precision. 
Other factors not related to a speaker's voice, however, do affect the estimation of the embedding vector representing the speaker \cite{takahashi14}.
As we know, speech utterances exhibit both extrinsic variability due to the background noises, channel distortions, as well as intrinsic variability including the physiological nature of the vocal apparatus and psychological states. 
It also has been pointed out that shorter speech utterances produce less reliable embeddings and result in poor ASV performance \cite{Zeinali20}. 
Therefore, it is essential to measure the precision, or equivalently, the uncertainty of the embeddings.

Few work has been done about the uncertainty of deep speaker embeddings.
For speaker diarization for which long recordings are typically cut into very short segments for clustering, a neural network has been used to predict the uncertainty of an x-vector \cite{Silnova20}, using the output of the statistics pooling layer of the original x-vector extractor. 
Then they are both input to an agglomerative hierarchical clustering (AHC) algorithm.
%
Xi-vector network \cite{lee21} with a posterior inference pooling has been proposed to integrate the Bayesian formulation of a linear Gaussian model to speaker-embedding neural networks. The precisions, however, are only used in the  calculation for posterior mean vectors from which xi-vectors are obtained and discarded afterwards.

Probabilistic linear discriminant analysis (PLDA) is a promising back-end for deep speaker embeddings \cite{wang22}.
The ability to handle uncertainty has been the cornerstone in the successful use of PLDA generative models \cite{dehak11,reynolds00,kenny08}. Thus, it is natural to propagate the uncertainty seen in the front-end to PLDA.
In this paper, we aim to incorporate the uncertainty that are calculated along the xi-vector extraction into the back-end PLDA scoring. Our contributions are:
\begin{inparaenum}[i)]
    \item we derive the formulation for the PLDA scoring with uncertainty propagation and
    \item propose to use a length scaling technique to replace the length normalization to enable its application to an uncertainty propagated back-end.
\end{inparaenum}

The paper is organized as follows.
Section 2 presents the PLDA with uncertainty propagation.
Section 3 derives the uncertainty in xi-vector space and presents a length scaling technique. 
Section 4 describes our experimental setup, results, and analyses. 
Section 5 summarizes our work. 

\vspace*{-1mm}
\section{Speaker Verification scoring with uncertainty}
\vspace*{-2mm}
Speaker verification can be accomplished by calculating the similarity between the two speaker embeddings 
corresponding to an enrollment and a test utterance. PLDA is a supervised parametric scoring method which is widely used in speaker recognition \cite{kenny10, Lee19}.
The propagation of speaker embeddings' uncertainty in PLDA has been addressed previously.  
I-vector posterior uncertainty, which is defined by the generative i-vector extraction model, is propagated to PLDA for the quality effect analysis caused by duration and phonetic variability \cite{kenny13,cumani14}.
In \cite{Silnova20} probabilistic embeddings, which consist of x-vectors and precisions, have been proposed to work with PLDA . The explicit form of the PLDA scoring log-likelihood ratio function, however, has not been presented.

\subsection{Uncertainty propagation in PLDA scoring (UP-PLDA)}
\label{ssec:up-plda}

Let $\phi_r$ be a $D$-dimensional speaker embedding vector of a speech utterance $r$ with a precision 
$\mathbf{\Phi}_{\mathrm{U},r}^{-1}$,
which measures the uncertainty in embedding extraction process.
We assume that the vector $\phi$ is generated from a linear Gaussian model \cite{bishop06}
\begin{equation}
\phi_r = \mu + \mathbf{Fy}+ \mathbf{U}_r \mathbf{x}+\sigma
\label{eq:gplda_up}
\end{equation}
with two latent variables: $\mathbf{y}\in \mathbb{R}^\mathit{d_y}$ and as the speaker variable
\begin{equation}
    \mathbf{y}\sim \mathcal{N}(\mathbf{0,I})
\end{equation}
and $\mathbf{x}\in \mathbb{R}^\mathit{d_x}$ as the variable to model the statistical noise in embedding extraction process
\begin{equation}
    \mathbf{x}\sim \mathcal{N}(\mathbf{0,I})
\end{equation}
 The vector $\mathbf{\mu}\in \mathbb{R}^\mathit{D}$ represents the global mean. The matrices $\bf{F}\in \mathbb{R}^{\mathit{D\times d}}$ is the speaker loading matrix, $\mathbf{U}_r$ is the lower-triangular Cholesky decomposition of the covariance $\mathbf{\Phi}_{\mathrm{U},r}$, which represents the uncertainty of the vector $\phi_r$, and $\sigma$ models the residual variances
\begin{equation}
    \sigma \sim \mathcal{N}(\mathbf{0,\Sigma})
\end{equation}
Integrating out the latent variables, we arrive at the following marginal density
\begin{equation}
    p({\phi_r})=\mathcal{N}\left(\phi|\mathbf{\mu}, \PhiB + \mathbf{\Phi}_{\mathrm{W},r} \right)
    \label{eq:marginal_density}
\end{equation}
where $\mathbf{\Phi}_\mathrm{B}$ is the between-speaker covariance matrix 
\begin{equation}
    \PhiB  = \mathbf{FF}\trans
    \label{eq:bw_up1}
\end{equation}
and $ \mathbf{\Phi}_{\mathrm{W},r}$ is an utterance-dependent within-speaker covariance matrix  
\begin{equation}
    \mathbf{\Phi}_{\mathrm {W},r} = \mathbf{\Sigma}+\mathbf{\Phi}_{\mathrm{U},r}
\label{eq:bw_up2}
\end{equation}
The total covariance correspondent to the utterance $r$ is the summation of the two covariance matrices 
\begin{equation}
    \mathbf{\Phi}_{\mathrm{tot},r} = \PhiB + \mathbf{\Phi}_{\mathrm {W},r}
    \label{eq:totr}
\end{equation}
and thus, is also utterance dependent.

In the testing phase, the log-likelihood ratio (LLR) between the enrollment ($\phi_\mathrm{e}$) and test ($\phi_\mathrm{t}$) embeddings is used to score how likely they are from the same speaker
\begin{equation}
s(\phi_\mathrm{e},\phi_\mathrm{t})=\log \frac{p(\phi_\mathrm{e},\phi_\mathrm{t})}{p(\phi_\mathrm{e})p(\phi_\mathrm{t})}
= \log \frac{\mathcal{N}(\phi_\mathrm{t}|\mathbf{M}_\mathrm{cond},\mathbf{\Sigma}_\mathrm{cond})}{\mathcal{N}(\phi_\mathrm{t}|\mathbf{M}_0, \mathbf{\Sigma}_0)}
\label{eq:llr_up}
\end{equation}
%
The probability density functions (pdf) are assumed to be conditioned on the model parameters and on $\phi_\mathrm{e}$. The predictive distribution in the numerator evaluated at $\phi_\mathrm{t}$ with mean and covariance equal to
\begin{equation}
\begin{aligned}
    \mathbf{M}_\mathrm{cond} &= \mu + \mathbf{F} \mathbf{M}_\mathrm{e} \\
    \mathbf{\Sigma}_\mathrm{cond} &= \mathbf{FL}_\mathrm{e}^{-1} \mathbf{F}\trans + \mathbf{\Phi}_{\mathrm{W,t}}
\end{aligned}
\label{eq:cond_up}
\end{equation}
which consists of estimating speaker vector posterior expectation $\mathbf{M}_\mathrm{e} $  and precision $\mathbf{L}_\mathrm{e}$
\begin{equation}
\begin{aligned}
    \mathbf{M}_\mathrm{e} &= \mathbf{L}_\mathrm{e}^{-1}\mathbf{F}\trans \mathbf{\Phi}_{\mathrm{W,e}}^{-1}\phi_\mathrm{e} \\
    \mathbf{L}_\mathrm{e} &= \mathbf{I}+ \mathbf{F}\trans \mathbf{\Phi}_{\mathrm{W,e}}^{-1}\mathbf{F}
\end{aligned}
\label{eq:mele}
\end{equation}
With the relations shown in (\ref{eq:bw_up1}-\ref{eq:bw_up2}) and Woodbury identity for inversion of matrices \cite{woodbury}, we use (\ref{eq:mele}) in \eqref{eq:cond_up} and obtain another form using the two covariance matrices
\begin{equation}
\begin{aligned}
   \mathbf{M}_\mathrm{cond}     &= \mu +\PhiB \Phitote \phi_\mathrm{e} \\
    \mathbf{\Sigma}_\mathrm{cond} &= \Phitott - \PhiB \Phitoteinv \PhiB
\end{aligned}
\label{eq:cond_up2}
\end{equation}
The denominator is also a normal pdf evaluated at $\phi_\mathrm{t}$ with parameters
\begin{equation}
\begin{aligned}
    \mathbf{M}_0 &= \mathbf{\mu} \\
    \mathbf{\Sigma}_0 &= \Phitott
\end{aligned}
\label{eq:m0sigma0}
\end{equation}
The posterior mean and covariance are set to their priors.

When uncertainty of speaker embeddings is not considered, i.e. the covariance $\mathbf{\Phi}_{\mathrm{U},r}$ is assumed to be zero, 
(\ref{eq:gplda_up}) becomes the traditional PLDA, and all embeddings share the same between-, within-, and total-speaker covariance matrices.
In the log-likelihood function (\ref{eq:llr_up}), 
the parameters of the predictive distribution in the numerator at $\phi_t$ are 
\begin{equation}
\begin{aligned}
    \mathbf{M}_\mathrm{cond} &= \mathbf{\mu} + \PhiB \Phitot \phi_\mathrm{e} \\
    \mathbf{\Sigma}_\mathrm{cond} &= \Phitot - \PhiB \Phitotinv \PhiB
\end{aligned}
\label{eq:cond1_ml2}
\end{equation}
and in the denominator $\mathbf{M}_0=\mathbf{\mu}$ and $\mathbf{\Sigma}_0 = \mathbf{\Phi}_\mathrm{tot}$.

Therefore, PLDA scoring LLR function with or without the uncertainty propagation has the same form. The difference is that with the uncertainty propagation, LLR has the within-speaker covariance that is dependent on the individual recording. They are adapted with an increase of uncertainty of the enrollment or test xi-vector estimate as shown in (\ref{eq:cond_up2}-\ref{eq:m0sigma0}).

\section{Derive embedding uncertainty from network }
\label{sec:pagestyle}
\subsection{Posterior inference in speaker embedding networks}
\label{ssec:xivec}
\begin{figure}[t]
\centering
\includegraphics[width=0.6\columnwidth]{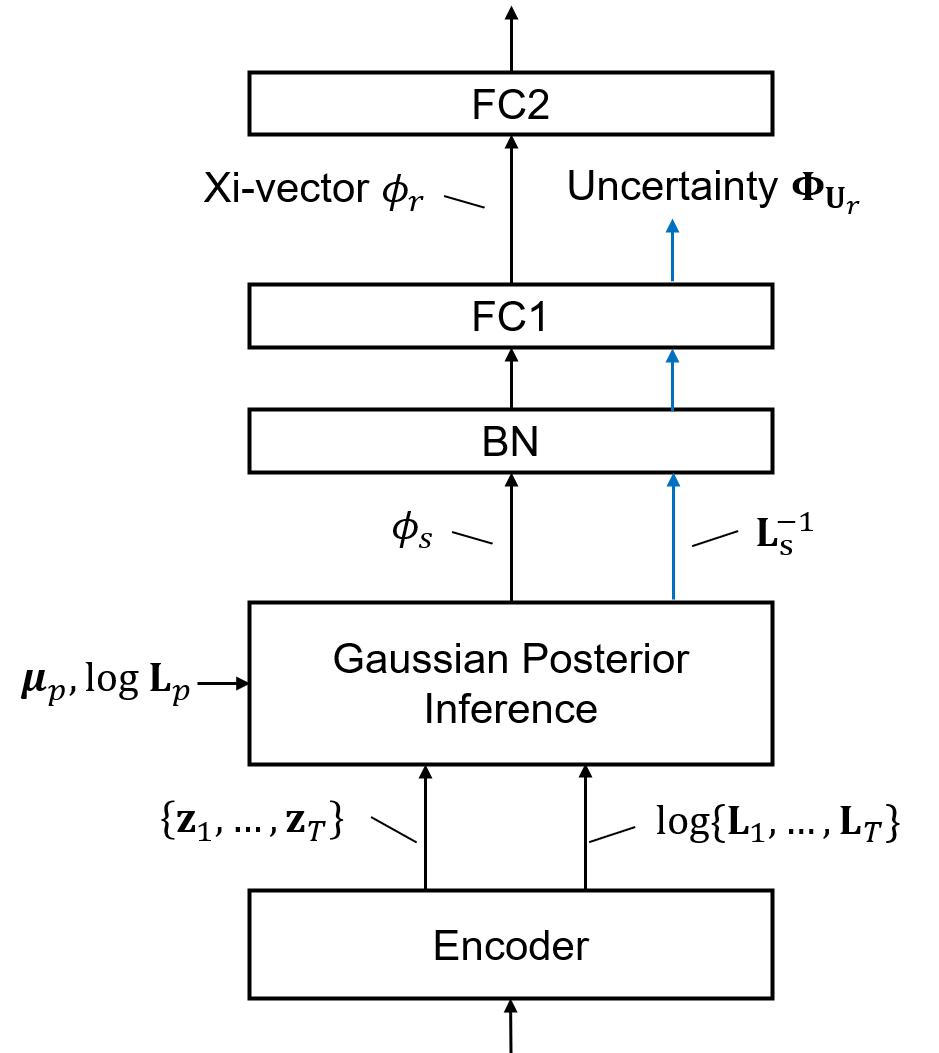}
\caption{{\it xi-vector network. Blue arrows are only for testing phase}}
\label{fig:network}
\end{figure}

Xi-vector \cite{lee21} was proposed to include a posterior inference pooling that integrates the Bayesian formulation of linear Gaussian model to speaker-embedding neural networks. It assumes that a linear Gaussian model is responsible for generating representations and characterizes the frame uncertainty with a covariance matrix associated with each estimate
\begin{equation}
    \mathbf{z}_t = \mathbf{h} + \epsilon_t
\end{equation}
where $\mathbf{h}$ is a latent speaker variable with the prior mean vector $\mu_p$ and covariance matrix $\mathbf{L}_p^ {-1}$
\begin{equation}
    \mathbf{h} \sim \mathcal{N}(\mu_\mathrm{p}, \mathbf{L}_\mathrm{p} ^ {-1}) 
\end{equation}
and $\epsilon_t$ represents uncertainty for the frame $t$
\begin{equation}  
        \epsilon_t \sim \mathcal{N}(\mathbf{0}, \mathbf{L}_t ^{-1})
\end{equation}
Given the input sequence and uncertainty estimate, posterior distribution of latent variable $\mathbf{h}$ is also Gaussian
\begin{equation}
    p(\mathbf{h}|\mathbf{z}_1, ..., \mathbf{z}_T, \mathbf{L}_1^{-1}, ... \mathbf{L}_T^{-1}) =\mathcal{N}(\mathbf{h}|\phi_\mathrm{s},\mathbf{L}_\mathrm{s} ^{-1})
\end{equation}
%
with a posterior mean $\phi_\mathrm{s}$ 
\begin{equation}
    \phi_\mathrm{s} = \mathbf{L}_\mathrm{s} ^{-1} (\sum_{t=1}^{T} \mathbf{L}_t \mathbf{z}_t + \mathbf{L}_\mathrm{p} \mathbf{z}_\mathrm{p})
    \label{eq:posterior_mean}
\end{equation}
and a precision matrix $\mathbf{L}_\mathrm{s}$
\begin{equation}
    \mathbf{L}_\mathrm{s} = \sum_{t=1}^{T} \mathbf{L}_t + \mathbf{L}_\mathrm{p}
    \label{eq:posterior_prec}
\end{equation}
%
Let $\mathbf{A}_t = \mathbf{L}_\mathrm{s}^{-1} \mathbf{L}_t$ for $t = 0,1,...T$, and the index $t = 0$ represents the prior such that $\mathbf{z}_0 =\mathbf{\mu}_\mathrm{p}$ and $\mathbf{L}_0 = \mathbf{L}_\mathrm{p}$, then the posteriors can be written 
\begin{align}
    \phi_\mathrm{s} = \sum_{t=1}^{T} \mathbf{A}_t \mathbf{z}_t &&
    \mathbf{L}_\mathrm{s} = \sum_{t=1}^{T} \mathbf{L}_t 
    \label{eq:phis_Ls}
\end{align}
%
After the posterior mean $\phi_\mathrm{s}$ is obtained in the Gaussian posterior inference layer (see Fig~\ref{fig:network}), it is followed by a batch normalization layer (BN) and a fully connected layer (FC1) from which xi-vectors are extracted.

\subsection{Derive xi-vector uncertainty}
The posterior precision $\mathbf{L}_s$ can be considered a measure of the uncertainty associated with the point estimate of latent variable $\mathbf{h}$. It is used in the mean estimation in \eqref{eq:posterior_mean} so as to extract xi-vector indirectly.  
It is, however, possible to propagate it to the space of xi-vector to represent its uncertainty associated with the embedding estimator besides a point estimate. In this way, the embedding is extended to be a distribution with the mean and variance $(\phi_r, \mathbf{\Phi}_{\mathrm{U},r} )$. 
To obtain this, we propagate the posterior covariance $\mathbf{L}^{-1}_s$ through the same layers accordingly (see Fig~\ref{fig:network}). The details are shown in Algorithm~\ref{alg:alg1}.

\subsection{Embedding length scaling (LS)}
\label{ssec:ls}
Length normalization (LN) together with whitening is a popular pre-processing technique to reinforce speaker embeddings to be more Gaussian distributed \cite{Garcia-Romero11}. It would be hard, however, to apply it to posterior covariance matrices because of its non-linearity. Therefore, we refine LN into an embedding length scaling (LS) technique 
\begin{equation}
    \phi_\mathrm{LS} = \left( \frac{d}{\phi\trans \mathbf{\Sigma} \phi} \right) ^{\frac{1}{2}}\phi
\label{eq:scalar}
\end{equation}
With the multiplication with $d^{\frac{1}{2}}$ in the numerator  to preserve the scaling of the original embedding space, we only need to apply LS to the enrollment and test embeddings in evaluations, not to PLDA training embeddings.
Note that when a total covariance matrix from the training data is used as $\mathbf{\Sigma}$, LS would be equivalent to LN but without retraining PLDA models.  
When the uncertainty of the embeddings is available, it can be propagated to LS (UP-LS) as well by including the embedding posterior covariance
\begin{equation}
    \mathbf{\Sigma}=\mathbf{\Phi}_{\mathrm{tot},r} 
\end{equation}


\begin{algorithm}[t]
\SetAlgoLined
  \textbf{Input} 
    Testing speech acoustic features $A=\{{a}_1, ..., {a}_T\}$ and 
    xi-vector network parameters  \\
  \textbf{Output} 
    xi-vector point estimate $\phi_r$ and uncertainty covariance $\mathbf{\Phi}_{\mathrm{U},r}$\\
Obtain frame-wise point estimate and its uncertainty \\
\hspace*{\algorithmicindent}$\{\mathbf{z}_1, ..., \mathbf{z}_T, \mathbf{L}_1^{-1}, ... \mathbf{L}_T^{-1}\} = f_\mathrm{enc}({{a}_1, ..., {a}_T)}$ \\
Obtain posterior mean vector and precision \\
\hspace*{\algorithmicindent}$\{\phi_\mathrm{s}, \mathbf{L}_\mathrm{s}\} = f_\mathrm{pool}(\mathbf{z}_1, ..., \mathbf{z}_T, \mathbf{L}_1^{-1}, ... \mathbf{L}_T^{-1})$ \\
Calculate xi-vector point estimate and posterior using BN layer parameters $(\mu_\mathrm{BN}, \sigma_\mathrm{BN})$ and FC1 layer parameters $(\mathbf{W},b)$\\
\hspace*{\algorithmicindent}$\phi_r = \mathbf{W}((\phi_\mathrm{s} - \mu_\mathrm{BN} ) \circ q_{\mathrm{BN}})+b$ \\
\hspace*{\algorithmicindent}$\mathbf{\Phi}_{\mathrm{U},r} =  \mathbf{W}(q_{\mathrm{BN}}{\trans} \mathbf{L}_\mathrm{s}^{-1} q_{\mathrm{BN}})\mathbf{W\trans}$ \\
where $q_{\mathrm{BN},i}=1/\sigma_{\mathrm{BN},i}$, $\circ$ is element-wise multiplication.
%
\caption{Derive xi-vector uncertainty from a network}
\label{alg:alg1}
\end{algorithm}

\section{experiments}
\label{sec:exp}
\newcommand{\bb}[1]{\parbox[t]{2mm}{\multirow{4}{*}{\rotatebox[origin=l]{270}{#1}}}}
{\renewcommand{\arraystretch}{1.1}
\begin{table*}[ht]
\centering
\caption{Performance of two ASV systems: ECAPA-TDNN x-vector and xi-vectors with PLDA on the four test sets: Vox1-O, Vox1-H, SITW core-core eval, and CNCeleb1-E. 
Two pre-processing techniques LN and LS are investigated. Results are shown as
EER(\%)/minDCF}    
\scalebox{0.9}{
\begin{tabular}{l|p{1.1cm}p{1.1cm}p{1.2cm}*{2}{|p{1.1cm}p{1.1cm}p{1.2cm}}|p{1.2cm}p{1.2cm}p{1.3cm}}
    \hline
    \multirow{2}{*}{} 
    & \multicolumn{3}{c|}{Vox1-O} 
    & \multicolumn{3}{c|}{Vox1-H}
    & \multicolumn{3}{c|}{SITW}
    & \multicolumn{3}{c}{CNCeleb} \\
    & - & LN & LS 
    & - & LN & LS 
    & - & LN & LS 
    & - & LN & LS    \\
    \hline                    
    \hline                    
    x
    & 1.44/0.184  &	1.23/0.173 & 1.23/0.173
    & 2.68/0.248  &	2.46/0.235 & 2.46/0.236
    & 1.80/0.180  & 1.67/0.172 & 1.69/0.172
    & 17.38/0.675 &	13.04/0.625 & 13.00/0.626
    \\
    xi
    & 1.49/0.166  & 1.23/0.143	& 1.23/0.141 
    & 2.76/0.247  &	2.44/0.236 &  2.43/0.235
    & 1.86/0.175  &	1.65/0.170 & 1.65/0.170
    & 17.32/0.675 & 12.64/0.629 & 12.63/0.630
    \\
    \hline   
\end{tabular}}
\label{tab:tab1}
\end{table*}}
{\renewcommand{\arraystretch}{1.1}
\begin{table*}[ht]
\centering
\caption{Performance of xi-vectors with PLDA and UP-PLDA. LS and UP-LS pre-processing techniques are evaluated. SITW* and CNCeleb* are the evaluations when mean adaptation is not used in centralization. Results are shown as EER/minCprimary}
\scalebox{0.9}{
\begin{tabular}{c| c c c c |c c}
    \hline
    Back-end  & Vox1-O   & Vox1-H  & SITW &  CNCeleb & SITW* & CNCeleb*
    \\
    \hline                    
    \hline                    
    PLDA            & 1.49/0.166 & 2.76/0.247 & 1.86/0.175 & 17.32/0.675 & 1.91/0.178 &17.99/0.689 \\
    LS PLDA         & 1.23/0.141 & 2.43/0.235 & 1.65/0.170 & 12.63/0.630 & 1.77/0.168 &15.04/0.676 \\
    UP-PLDA         & \textbf{0.99}/0.129 & \textbf{2.13}/0.231 & 1.77/\textbf{0.167} & 10.81/0.652 & 1.80/0.170 &13.18/0.714 \\
    UP-LS UP-PLDA   & 1.01/\textbf{0.124} & 2.18/\textbf{0.224} & \textbf{1.59/0.167} & \textbf{10.16/0.608} & \textbf{1.59/0.166} & \textbf{11.57/0.686} \\   
\hline                    
   \end{tabular}}
    \label{tab:tab2}
    \label{exp2}
\end{table*}}

\subsection{Experimental settings} 
The experiments were conducted on the VoxCeleb \cite{Nagrani17},
the Speaker in the Wild (SITW) core-core  {\it eval} \cite{McLaren16}, and the CNCeleb1 dataset \cite{fan20}. For VoxCeleb1, we exploited the original test set Vox1-O and the hard test set Vox1-H.
The front-end networks were trained using the original segments of VoxCeleb2 dataset \cite{Chung18} with augmentations following the settings in \cite{wang22}. 
The same training dataset without augmentation was used to train back-ends after all VoxCeleb2 segments belonging to the same session were concatenated.
For SITW and CNCeleb evaluations, SITW core-core {\it dev} set and CNCeleb1-T dataset were used, respectively, as the development sets for the mean normalization. 

We used x-vector and xi-vector networks with an ECAPA-TDNN \cite{Desplanques20} backbone and optimized them with AAM-Softmax cross-entropy loss \cite{deng18}. We used their standard forms. In the x-vector network, both weighted means and standard deviations from an attentive statistics pooling layer were propagated, while in xi-vector network only the posterior mean from the Gaussian posterior inference layer was propagated. 
Both types of embeddings have 192 dimensions. The posterior covariance matrices in the xi-vector network were assumed diagonal \cite{lee21}, and the prior mean and covariance were initialized as $(0,I)$. In the testing phase , we obtained the corresponding posterior covariance following Algorithm~\ref{alg:alg1}.

For the back-end, the advantage of using the diagonalized within-speaker covariance matrix in PLDA has been proved \cite{wang22}. We further diagonalized the between-speaker covariance matrix for the computational efficiency. We used the total covariance of the training data in LS for PLDA back-ends and that plus testing utterance's posterior covariance in UP-LS for UP-PLDA back-end.  Raw embeddings were used in PLDA training in the systems where LS was applied to testing data. 
We used the SpeechBrain open-source toolkit \cite{sb21} for the front-end implementations and embedding extractions. The input of the neural networks were 80-dimensional filter-bank features.
Results are reported in terms of equal error rate (EER) and the minimum normalized detection cost function (MinDCF) at $P_\mathrm{target}= 10^{-2}$ and $C_\mathrm{FA} = C_\mathrm{Miss} = 1$.

\begin{figure}[t]
\centering
\includegraphics[width=0.9\columnwidth]{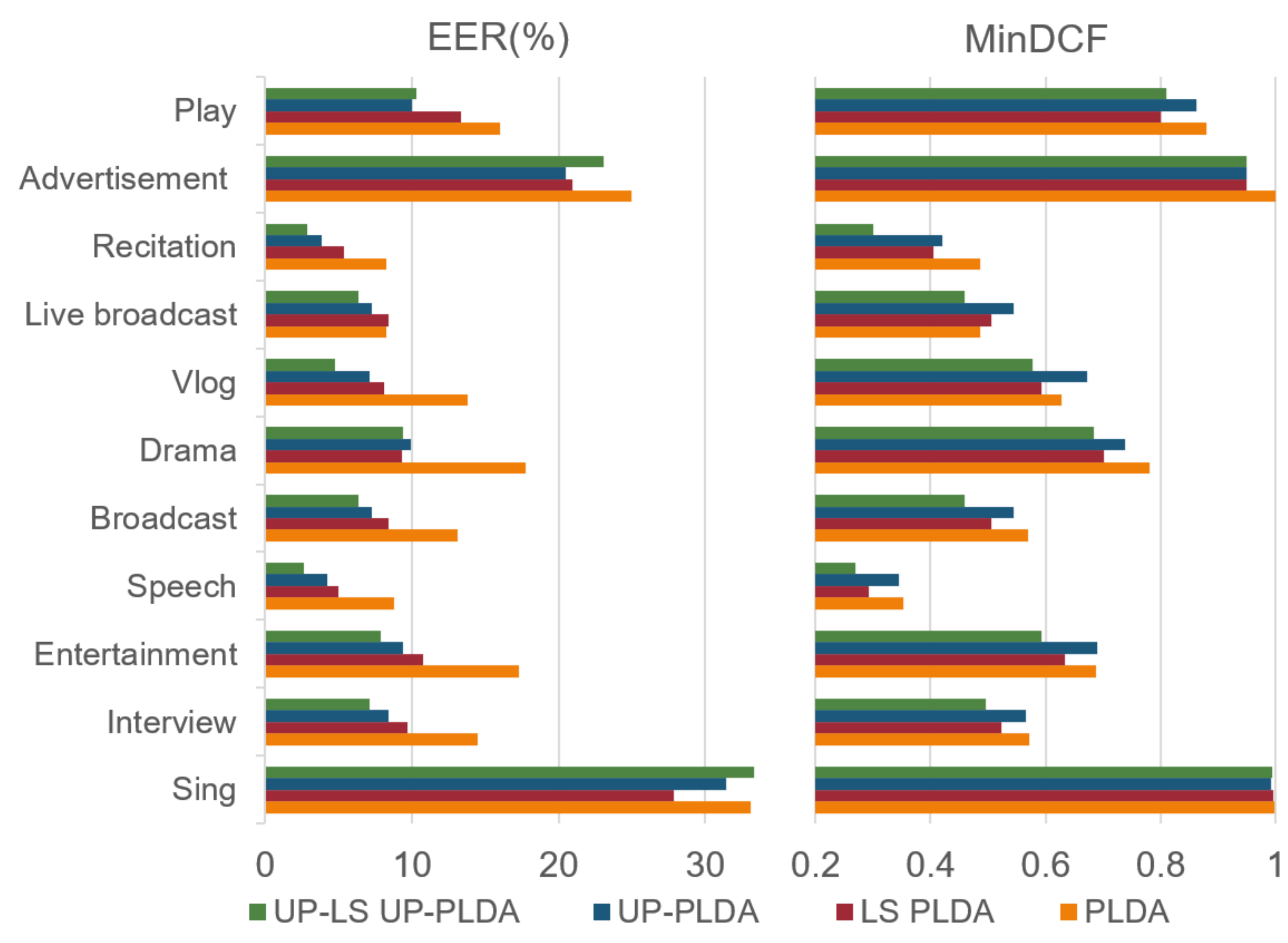}
\caption{{\it Evaluation of length scaling (LS) and uncertainty propagation (UP) in different genres in CNCeleb1-E set}}
\label{fig:cn}
\end{figure}

\subsection{Experimental results}
We first compare LS and LN pre-processing techniques in both x-vector and xi-vector PLDA systems. As shown in Tab.~\ref{tab:tab1}, the uses of LN and LS give almost the same results, and they are consistently better than those without any pre-processing. It proves the equivalency between LN and LS when using the total covariance matrix as mentioned in \ref{ssec:ls}. Thus, we next apply LS in the back-end of PLDA with uncertainty.
The advantage of xi-vectors is not obvious over x-vectors. We argue that features with different time-scales derived from the dense connection in ECAPA-TDNN may cause confusion to the xi-vector posterior inference process.

Next we evaluate the uncertainty propagation in LS pre-processing and in PLDA, referred as UP-LS and UP-PLDA, respectively.
Table~\ref{tab:tab2} shows a significant improvement due to the explicit use of xi-vector uncertainty in UP-PLDA scoring (line 3) over the system without it (line 1). 
Such observations are consistent in all four evaluation sets.
It indicates that the explicit use of uncertainty in the back-end is more effective than its implicit use in xi-vector estimation.
For the Vox1 and SITW evaluations, UP-PLDA yields a greater improvement in both EER and minDCF than that LS does, while for CNCeleb evaluation, UP-PLDA gives a better EER but a slight increase in minDCF. Further application of UP-LS pre-processing to xi-vectors, for UP-PLDA back-end, improved ASV performance in the SITW and CNCeleb evaluation sets in which certain mismatches exist in domains.   
Overall, the use of LS pre-processing and uncertainty propagation in PLDA together achieved the best performance, with $14.5\%$--41.3$\%$ and $4.6\%$--$25.3\%$ reductions, respectively, in EER and minDCF.
For the SITW and CNCeleb sets, we also show in Tab~\ref{tab:tab2} the performance using the mean of the training data for the embedding centralization. The comparison shows a clear advantage of using the mean adaptation in domain mismatched evaluations.

The CNCeleb evaluations give large values in EER and minDCF in all the systems (see Tab~\ref{tab:tab1} and Tab~\ref{tab:tab2}) due to its severe conditions. We next investigate the UP-PLDA and LS effects in each genre, as shown in Fig~\ref{fig:cn}.  Despite the difference in language and speaking styles between CNCeleb set and the training data VoxCeleb, the observations of the improvement due to the use of UP-PLDA and LS pre-processing, in most of the genres, are consistent with the overall results in Tab~\ref{tab:tab2}. 


\section{Summary}
This paper has revisited uncertainty propagation in PLDA. Based on the xi-vector framework, we derive a posterior covariance matrix from frame-wise precisions, to measures the uncertainty of speaker embeddings. We propose a log-likelihood ratio function for the PLDA scoring with the propagation of embedding uncertainty. At last, we propose to replace the length normalization pre-processing technique with a length scaling technique for the application of uncertainty propagation in the back-end. 
Experimental results on the VoxCeleb-1, SITW core-core eval sets as well as the domain-mismatched CNCeleb1 set show the effectiveness of the two techniques with $14.5\%$--41.3$\%$ and $4.6\%$--$25.3\%$ reductions, respectively, in EER and minDCF. In future, we will investigate the networks with different backbones and the use of uncertainty in domain adaptation. 

\section{Acknowledgements}
This project is supported by the Agency for Science, Technology and Research (A$^\star$STAR), Singapore, through its Council Research Fund (Project No. CR-2021-005).
\bibliographystyle{IEEEbib}
\bibliography{strings,refs}

\end{document}